\documentclass[reprint,superscriptaddress,amsmath,amssymb,aps,prx]{revtex4-2}

\usepackage{siunitx}
\usepackage{graphicx}
\usepackage{bm}

\usepackage{physics}
\usepackage[dvipsnames]{xcolor}
\usepackage[
  colorlinks=true,
linkcolor=Blue,
  citecolor = Blue,
  urlcolor = Blue,
  unicode
  ]{hyperref}
\usepackage{natbib}
\usepackage[nameinlink]{cleveref}
\crefname{equation}{Eq.}{Eqs.}
\Crefname{equation}{Equation}{Equations}
\crefname{figure}{Fig.}{Figs.}

\Crefname{figure}{Figure}{Figures}

\usepackage{ulem}

\begin{document}

\title{Entanglement-assisted multiparameter estimation with a solid-state quantum sensor}

\author{Takuya Isogawa} 
\thanks{These authors contributed equally.}
\affiliation{Research Laboratory of Electronics, Massachusetts Institute of Technology, Cambridge, MA 02139, USA}
\affiliation{Department of Nuclear Science and Engineering, Massachusetts Institute of Technology, Cambridge, MA 02139, USA}

\author{Guoqing Wang} 
\thanks{These authors contributed equally.}
\affiliation{Research Laboratory of Electronics, Massachusetts Institute of Technology, Cambridge, MA 02139, USA}
\affiliation{Department of Physics, Massachusetts Institute of Technology, Cambridge, MA 02139, USA}

\author{Boning Li} 
\thanks{These authors contributed equally.}
\affiliation{Research Laboratory of Electronics, Massachusetts Institute of Technology, Cambridge, MA 02139, USA}
\affiliation{Department of Physics, Massachusetts Institute of Technology, Cambridge, MA 02139, USA}

\author{Zhiyao Hu} 
\affiliation{Research Laboratory of Electronics, Massachusetts Institute of Technology, Cambridge, MA 02139, USA}
\affiliation{Pritzker School of Molecular Engineering, The University of Chicago, Chicago, IL 60637, USA}

\author{Shunsuke Nishimura} 
\affiliation{Research Laboratory of Electronics, Massachusetts Institute of Technology, Cambridge, MA 02139, USA}
\affiliation{Department of Nuclear Science and Engineering, Massachusetts Institute of Technology, Cambridge, MA 02139, USA}
\affiliation{Department of Physics, The University of Tokyo, Bunkyo-ku, Tokyo, 113-0033, Japan}

\author{Ayumi Kanamoto} 
\affiliation{Research Laboratory of Electronics, Massachusetts Institute of Technology, Cambridge, MA 02139, USA}
\affiliation{Department of Nuclear Science and Engineering, Massachusetts Institute of Technology, Cambridge, MA 02139, USA}
\affiliation{Department of Electrical and Electronic Engineering, Institute of Science Tokyo, Meguro,
Tokyo 152-8550, Japan}

\author{Haidong Yuan} 
\email{hdyuan@mae.cuhk.edu.hk}
\affiliation{
   Department of Mechanical and Automation Engineering, The Chinese University of Hong Kong, Shatin, Hong Kong}
\affiliation{The Hong Kong Institute of Quantum Information Science and Technology, The Chinese University of Hong Kong, Shatin, Hong Kong SAR, China}
\affiliation{State Key Laboratory of Quantum Information Technologies and Materials, The Chinese University of Hong Kong, Shatin, Hong Kong SAR, China}   
\author{Paola Cappellaro} 
\email{pcappell@mit.edu}
\affiliation{Research Laboratory of Electronics, Massachusetts Institute of Technology, Cambridge, MA 02139, USA}
\affiliation{Department of Nuclear Science and Engineering, Massachusetts Institute of Technology, Cambridge, MA 02139, USA}
\affiliation{Department of Physics, Massachusetts Institute of Technology, Cambridge, MA 02139, USA}

\date{\today}

\begin{abstract}
Quantum multiparameter estimation promises to extend quantum advantage to the simultaneous high-precision measurements of multiple physical quantities. However, realizing this capability in practical quantum sensors under realistic conditions remains challenging due to intrinsic system imperfections. Here, we experimentally demonstrate multiparameter estimation using a nitrogen-vacancy (NV) center in diamond, a widely adopted solid-state quantum sensor. Leveraging electronic-nuclear spin entanglement and optimized Bell state measurement at room temperature, we simultaneously estimate the amplitude, detuning, and phase of a microwave drive from a single measurement sequence. Despite practical constraints, our results achieve linear sensitivity scaling for all parameters with respect to interrogation time. This work bridges the gap between foundational quantum estimation theory and real-world quantum sensing, opening pathways toward enhanced multiparameter quantum sensors suitable for diverse scientific and technological applications.
\end{abstract}

\maketitle

\begin{figure*}
 
\centering
\includegraphics[width=\textwidth]{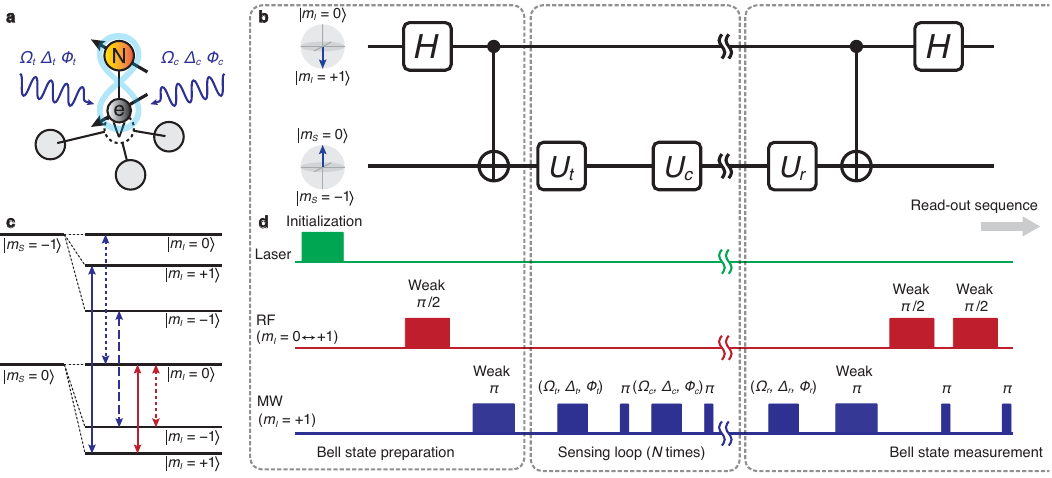}
\caption{\textbf{Principles of multiparameter estimation based on a single NV center in diamond.} \textbf{a}, A single NV center is used to estimate the three target parameters of the microwave field. The electronic spin is used as a sensor qubit, while the nitrogen nuclear spin serves as an ancilla qubit.  Entanglement between the two qubits and quantum control via microwave fields are used to enhance the sensitivity. \textbf{b}, Quantum circuit diagram of the experimental demonstration. It consists of three parts: Bell state preparation, sensing loop, and Bell state measurement. \textbf{c}, Relevant energy levels of the NV center. \textbf{d}, Optical, radio-frequency (RF), and microwave (MW) pulse sequence used for implementing the quantum circuit shown in (\textbf{b}). The RF field drives the transition between \( m_I = 0 \) and \( m_I = +1 \) in the \( m_S = 0 \) manifold. The microwave field drives the transition between \( m_S = 0 \) and \( m_S = -1\) in the \( m_I = +1 \) manifold. To prepare the Bell state, the system is first initialized to the \( m_S = 0, m_I = +1 \) state with a long laser pulse. A \(\pi/2\) pulse is then applied to the nuclear spin, followed by a selective (weak) \(\pi\) pulse on the electronic spin. 
The system is subjected to the target microwave field and a control microwave field, which is flanked by two \(\pi\) pulses to decouple the interaction between the sensor and ancilla qubits. This sensing loop is repeated \(N\) times. Subsequently, a microwave pulse for the rotation of the measurement basis, a selective microwave \(\pi\) pulse and a non-selective RF \(\pi/2\) pulse are applied. To implement a non-selective RF \(\pi\)/2 gate on the nuclear spin we embed a non-selective microwave \(\pi\) pulse into two consecutive RF \(\pi\)/2 pulses. Although omitted from the figure, a selective microwave \(\pi\) pulse is applied at the end to switch the basis.} 
\label{fig:fig1}
\end{figure*} 

Quantum sensors have found a broad spectrum of applications ranging from biological measurements to investigations of condensed matter phenomena~\cite{Giovannetti2011,RevModPhys.89.035002,Aslam2023,Casola2018}. While the majority of these applications have focused on the estimation of a single parameter, there is a growing interest in quantum multiparameter estimation~\cite{PhysRevA.75.022326,PhysRevA.69.022303,Imai_2007,PhysRevLett.111.070403,PhysRevLett.116.030801,Szczykulska03072016,PhysRevLett.125.020501,PhysRevA.94.052108,Liu_2020,Demkowicz-Dobrzański_2020,vasilyev2024optimalmultiparametermetrologyquantum,PhysRevLett.117.160801,Hou2016,PhysRevLett.112.103604,Taylor2013,PhysRevLett.124.060502,Polino:19,Hou2018,Roccia_2018,Ciampini2016,PhysRevA.102.022602,Zhou:15,doi:10.1126/sciadv.abd2986, PhysRevLett.126.070503,hu2024controlincompatibilitymultiparameterquantum,zhang2024distributedmultiparameterquantummetrology}, 
reflecting both the fundamental challenges linked to non-commutativity and the practical need to fully characterize a complex system or environment by simultaneously measuring multiple physical parameters.
 By harnessing quantum resources such as entanglement, quantum multiparameter estimation can enable the simultaneous determination of multiple parameters without incurring significant tradeoffs in precision. This capability is particularly powerful in real-world applications, where reducing the number of sensors or measurement protocols translates into savings in time and resources. Despite  proof-of-principle experiments conducted using photons~\cite{Hou2016,PhysRevLett.112.103604,Taylor2013,PhysRevLett.124.060502,Polino:19,Hou2018,Roccia_2018,Ciampini2016,Zhou:15,doi:10.1126/sciadv.abd2986, PhysRevLett.126.070503} and superconducting qubits~\cite{zhang2024distributedmultiparameterquantummetrology}, a systematic study of the impact of practical experimental limitations is still lacking. In particular, achieving quantum multiparameter estimation in realistic settings --using widely adopted quantum sensors and operating under practical experimental constraints-- is  crucial for translating theoretical advantages of quantum metrology into tangible benefits for sensing technologies.

 In this work, we demonstrate quantum multiparameter estimation using a solid-state quantum sensor, specifically the nitrogen vacancy (NV) center in diamond (Fig.~\ref{fig:fig1}a). Previous sensing studies using NV centers have demonstrated measurements of DC and AC magnetic field vector components, as well as the amplitude, frequency, and phase of AC magnetic fields~\cite{Maze2008,Taylor2008,10.1063/1.3337096,10.1063/1.3385689,Pham_2011,Tetienne_2012,Wang2015,https://doi.org/10.1002/adom.201600039,PhysRevLett.122.100501,PhysRevApplied.10.044039,RevModPhys.92.015004,PhysRevX.12.021061,Wang2021,PhysRevA.107.062423}. NV centers have also been used to detect electric fields, temperature, and pressure~\cite{Dolde2011,Kucsko2013,PhysRevB.100.174103,Huxter2023,doi:10.1126/science.aaw4352}. However, all of these demonstrations relied on combinations of separate measurement sequences designed for single-parameter estimation. Consequently, none were based on quantum multiparameter estimation, where multiple parameters are simultaneously extracted from a single measurement sequence by analyzing the resulting probability distribution. In our implementation, we introduce a three-parameter sensing protocol leveraging Bell states within a two-qubit system composed of a sensor qubit and an ancilla qubit (Fig.\ref{fig:fig1}b). This configuration serves as a paradigmatic model for quantum multiparameter estimation, where the sensor qubit interacts directly with the vector field \( \vec{B} \), while the ancilla qubit remains decoupled from it. Such multiparameter estimation schemes have garnered significant recent interest\cite{PhysRevA.69.022303,PhysRevLett.117.160801,doi:10.1126/sciadv.abd2986,zhang2024distributedmultiparameterquantummetrology}. In previous experimental efforts employing photons~\cite{doi:10.1126/sciadv.abd2986} and superconducting qubits~\cite{zhang2024distributedmultiparameterquantummetrology}, the interactions with target fields were simulated through decomposed gate sequences. In contrast, our experiment directly estimates three physically meaningful parameters from a real microwave field—its amplitude, frequency, and phase.

\section{\label{sec:case_study}Three-parameter sensing}
We employ the electronic-nuclear spin system of the $^{14}$NV center~\cite{Hirose2016,Hernández-Gómez2024}. Its Hamiltonian is given by
\(
\widetilde H_{SI} = D S_z^2 + \vec{B}(t)\cdot(\gamma_e \vec{S}+ \gamma_n \vec{I}) + Q I_z^2  + A S_z I_z,
\)
where $D = (2\pi)\times 2.87\,\mathrm{GHz}$ is the zero-field splitting, $Q = -(2\pi)\times 4.95$~MHz~is the nuclear quadrupole moment, $\gamma_{e}\approx (2\pi)\times 2.8\,\mathrm{MHz/G}$, $\gamma_{n}\approx-(2\pi)\times 0.31\,\mathrm{kHz/G}$ denote the gyromagnetic ratios of the electronic and nuclear spins, $A = -(2\pi)\times2.16$ MHz is the hyperfine coupling constant, $\vec S$ and $\vec I$ are the 
spin-1 operators of the electronic and nuclear spins, respectively. The total magnetic field $\vec B(t)$ includes a static component $B^0_z= 357\,$G  along the NV axis and AC control fields and the target field.
We define two qubits within the subspace spanned by the four levels $\{|m_S, m_I\rangle = |0,+1\rangle, |0,0\rangle, |-1,+1\rangle, |-1,0\rangle\}$, while all six levels shown in Fig.~\ref{fig:fig1}c are used during the Bell state measurement. At room temperature and under this moderate static field, the readout is restricted to ensemble-averaged fluorescence detection of the electronic spin (i.e., no single-shot readout). We employ microwave fields to control the electronic spin and radio-frequency (RF) fields to drive the nuclear spin; by adjusting the amplitude, frequency, and phase of these fields, we implement the requisite gate operations (Fig.\ref{fig:fig1}d). To analyze the system’s dynamics, we move to the rotating frame defined by $H_0=(D-\gamma_e B^0_z)S_z+(Q+\gamma_nB^0_z)I_z$. Expressing the spin-1 operators in terms of effective spin-1/2 operators yields the two-qubit interaction Hamiltonian,
$H_\textrm{int}=A (-\sigma_z^e + \sigma_z^n - \sigma_z^e \sigma_z^n)/4$.

The multiparameter sensing task is to determine three parameters \(\theta =(\Omega_t,\Delta_t,\Phi_t)\) associated with a microwave drive. While the field is time dependent in the frame set by $H_0$, we can express it in an appropriate interaction picture as $H(\Omega,\Delta,\Phi) = \frac{\Delta}{2}\sigma^{e}_z+\frac{\Omega}{2}(\cos{(\Phi)}\sigma^{e}_x - \sin{(\Phi)}\sigma^{e}_y)$, where $\Omega$ is the Rabi frequency, $\Delta$ is the detuning, and $\Phi$ is the phase of the microwave.
The system evolution under the target field is then described by $U_t = \exp(+i\Delta_t  \sigma^{e}_z T/2)\exp(-i (H(\Omega_t,\Delta_t,\Phi_t)+H_{\text{int}})T)$. Unlike the ideal unitary evolution acting only on the electronic spin, the presence of hyperfine interaction in $H_{\text{int}}$ induces more complex two-qubit dynamics, which we address in a subsequent section.

The route to achieving high-sensitivity multiparameter sensing includes (i) preparing an optimal initial state, (ii) choosing an optimal measurement, and (iii) tailoring the time evolution to optimally extract all three parameters. In the ideal case to estimate the vector field $\vec{B}$, it is known from the quantum Fisher information (QFI) arguments that the Bell states and the Bell state measurement are optimal probe state and measurement, respectively, and that one can achieve linear scaling for all three parameters in the interrogation time $T$ for small fields $|\vec{B}|\to0$, $\delta B_i\propto1/T$ for $i \in\{x, y, z\}$.~\cite{PhysRevLett.117.160801, doi:10.1126/sciadv.abd2986}. Achieving such scaling is nontrivial, both in the context of multiparameter estimation and even in single-parameter estimation when the Hamiltonian is not in the multiplicative
form of the parameter \(H(x)=x\,G\)~\cite{PhysRevLett.115.110401}. Moreover, in realistic solid-state quantum sensors based on spin systems such as the one employed in our experiment, several non-idealities arise: (i) the electron and nuclear spins are always coupled via the hyperfine term \(\tfrac{A}{4}\sigma_z^e\sigma_z^n\); (ii) only ensemble-averaged fluorescence from the electronic spin is detected, precluding single-shot two-qubit readout; and (iii) measurements are ultimately photon-shot-noise limited. These significant differences from an ideal system make it a highly non-trivial task to find optimal sensing strategies for practical implementation. In the following, we carefully analyze these effects and develop optimal control and Bell-measurement protocols to recover near-ideal performance.

\section{\label{sec:control_strategy}Sequential control scheme}

To achieve linear sensitivity scaling for all three parameters, we build upon the sequential control scheme introduced in Ref.~\cite{PhysRevLett.117.160801}. By applying the control gate \( U_c \approx U_t^\dagger \) immediately after the target gate \(U_t\) and repeating this sequence \( N \) times, an effective bias field can be introduced, allowing detection in regions where the target magnetic field is small, which is equivalent to the condition $|\vec{B}|\to0$ in the ideal vector field model. 

To implement \( U_t^\dagger \), we need to invert the sign of the total Hamiltonian \(H_{\text{tot}} = H(\Omega_t, \Delta_t, \Phi_t) + H_{\text{int}}.\) Although the first term can be easily inverted by simply adjusting \( (\Omega, \Delta, \Phi) \), the always-on ancilla coupling $H_{\text{int}}$ cannot simply be switched off. The effect of the \(\sigma_z^{n}\) term in \( H_{\text{int}} \) can be easily canceled by changing the phase of the second \(\pi/2\) pulse applied to the nuclear spin (See Supplementary Section 6). Furthermore, we address the terms \( \sigma^{e}_z \sigma^{n}_z \) and \(\sigma_z^{e}\) by introducing a dynamical decoupling approach. The idea is to apply spin-flip $\pi$ pulses to the electronic spin to effectively reverse the sign of the interaction term \( \sigma^{e}_z \sigma^{n}_z \), leveraging the identity \(\sigma_x \sigma_z \sigma_x = -\sigma_z\).
For simplicity, we first consider the near-resonant case $\Delta_t \approx 0$. Setting $\Omega_c = \Omega_t$, $\Delta_c = \Delta_t$, and $\Phi_c = \pi - \Phi_t$ ensures $U_{\pi} \, e^{-i H_c T_t} \, U_{\pi} = U_t^\dagger$, where $U_\pi = \sigma_x$ and $H_c=H(\Omega_c, \Delta_c, \Phi_c) + H_{\text{int}}$. For nonzero detuning $\Delta_t \neq 0$, an additional frame-change factor $\exp\bigl(+i \Delta_t \,\sigma^{e}_z \,T/2\bigr)$ emerges. To cancel this extra phase, we further refine the control scheme by setting the phase of the first electronic spin $\pi$-pulse as $2\pi \Delta_t T$, thereby achieving optimal control even under more general target field conditions.

\section{\label{sec:Bell_state_measurement}Bell state measurement}
\begin{figure*}
\centering
\includegraphics[width=\textwidth]{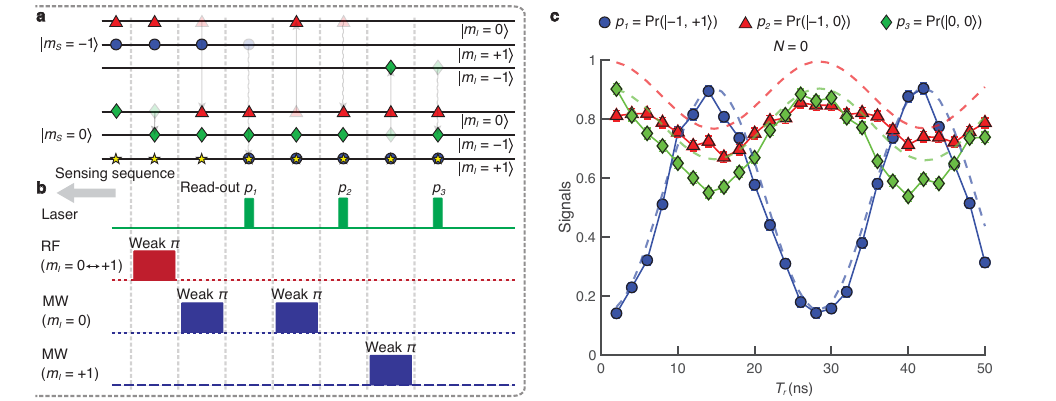}
\caption{\textbf{Readout sequence and experimental demonstration of Bell state measurement.} \textbf{a}, Energy-level diagram illustrating the read-out sequence. After applying the disentangling gates shown in Fig.~\ref{fig:fig1}, the Bell state measurement reduces to measuring the population of the $\ket{-1,+1}$ (\(p_1,\) blue circle), $\ket{-1,0}$ (\(p_2,\) red triangle) and $\ket{0,0}$ (\(p_3,\) green diamond) states.
\textbf{b}, Optical, radio-frequency (RF), and microwave (MW) pulse sequence used for the read-out shown in (\textbf{a}). Initially, a nuclear $\pi$ pulse between the $m_I = 0$ and $m_I = -1$ states shelves population $p_3$ into the unused $\ket{m_S=0, m_I=-1}$ state. Next, a selective (weak) microwave $\pi$ pulse conditioned on $m_I = 0$ transfers population $p_2$ to the $\ket{0,0}$ state. A short laser pulse measures the electronic spin, yielding $p_1$ without affecting the nuclear spin. Although this measurement erases the distinction between $p_1$ and $p_4$, the shelving enables extraction of both $p_2$ and $p_3$, which are sequentially mapped onto the electronic spin for optical readout. \textbf{c}, Example of the Bell state measurement as a function of the pulse duration for the rotation gate $T_r$ for fixed amplitude $\Omega_r=(2\pi)\times21\ \text{MHz}$, detuning $\Delta_r=(2\pi)\times30\ \text{MHz}$, and phase $\Phi_r= 30^\circ\ $. We plot the normalized signals without any sensing loop ($N=0$). To cancel the frame-change factor $\exp\bigl(+i \Delta_t \,\sigma^{e}_z \,T/2\bigr)$, we set the phase of the following $\pi$-pulse as $2\pi \Delta_r T_r$. The dashed lines represent a model that accounts for state preparation and measurement (SPAM) errors due to non-ideal behavior of nuclear spin states under laser irradiation; see Supplementary Section 5 for details. Error bars represent the standard deviation of the signal determined through error propagation from the photoluminescence intensity. Each displayed point is a result of $n=4\times10^6$ averages of sequences.}
\label{fig:BSM}
\end{figure*}

In conventional single-parameter sensing, each measurement sequence yields only one outcome, limiting the extracted information to a single parameter. By entangling the sensor with an ancilla qubit, we effectively expand the measurement basis to two qubits, enabling simultaneous estimation of multiple parameters. A Bell state measurement on this entangled system, then, provides all three target parameters from a single measurement sequence, thus realizing genuine quantum multiparameter sensing.

Bell state measurements of the electronic-nuclear spin system associated with the NV center have been demonstrated in conditions that are not always compatible with the most general sensing tasks, in particular at cryogenic temperatures or under high magnetic fields~\cite{Robledo2011,doi:10.1126/science.1253512, Kamimaki2023}. Here we achieve averaged readout of the Bell basis at room temperature and under a low static DC field by utilizing the three-level system of the nitrogen-14 nuclear spin (see Fig.~\ref{fig:BSM}). The key insight is that we can extract information about both the electron and nuclear spin states by performing three electronic spin readouts, while the non-qubit level of the nuclear spin acts as a (classical) memory. Specifically, we first transfer the \( m_S = 0, m_I = 0 \) state to the \( m_S = 0, m_I = -1 \) state, which lies outside the qubit subspace and remains unpopulated, effectively "shelving" the state. This allows for determining the three populations $p_i,\,i\in\{1,2,3\}$ after CNOT and Hadamard gates are applied to transform the state into the computational basis, as illustrated in Fig.~\ref{fig:BSM}a,b.

This readout sequence enables the measurement of three signals as shown in Fig.~\ref{fig:BSM}c, which presents the measurement results obtained by rotating the Bell basis without performing the sensing loop, i.e., $N=0$. The dashed lines indicate a simple model that accounts for state preparation and measurement (SPAM) errors due to the non-ideal behavior of nuclear spin states under laser irradiation. As illustrated in Fig.~\ref{fig:fig1}, the Bell state is prepared by applying a $\pi/2$ pulse ($\exp(+i\pi\sigma_y/4)$) on the nuclear spin and a selective $\pi$ pulse on the electronic spin. Due to optical pumping, while the electronic spin becomes nearly fully polarized, the nuclear spin achieves only partial polarization, quantified by $P=0.85$. This results in a mixed initial state
\(
\rho = P\,|0,+1\rangle\langle0,+1| + (1-P)\,|0,0\rangle\langle0,0|,
\)
which, after the entangling gates, evolves into a classical mixture of Bell states:
\(
\rho = P\,|\Phi_+\rangle\langle\Phi_+| + (1-P)\,|\Phi_-\rangle\langle\Phi_-|,
\) where \(|\Phi_{\pm}\rangle=(|0,0\rangle \pm |-1,+1\rangle)/\sqrt{2}\). For the measurement error, we incorporate a simple linear model to capture the effects of nuclear spin population transfer during short laser pulses used for electron readout. The cumulative effect of "leakage" on the state populations is described by the transformation: \(p'=M p\), where \(M\) is a matrix parametrized by the leakage rates \(\zeta=0.20,\gamma=0.15,\) and \(\eta=0.025\). The parameters \(P\), \(\zeta\), \(\gamma\), and \(\eta\) are determined experimentally, as detailed in Supplementary Section~4.
Notably, with no free parameters, this model accurately reproduces the experimental data, particularly the signal amplitudes, without accounting for additional experimental imperfections such as pulse errors. As will be discussed later, the signal amplitude directly impacts the sensor's sensitivity.

\section{Sensitivity analysis}

We quantify the sensor’s performance by extracting the parameter covariance matrix \(\Sigma_{\hat{\theta}}\) from the measurement signals. For small variations in \(\theta\), the diagonal elements of \(\Sigma_{\hat{\theta}} \;=\; J^{-1} \,\Sigma_{\hat{p}}\, (J^{-1})^\intercal\) yield the sensitivities, where \(\Sigma_{\hat{p}}\) is the covariance matrix of the measured signals, and \(J_{ij}=\partial p_i/\partial \theta_j\) is the Jacobian matrix relating changes in \(\theta\) to the measured signals. Under shot-noise-limited detection with uncertainty \(\sigma\), we have \(\Sigma_{\hat{p}} = \sigma^2 \mathbf{I}\). A similar analysis applies in the single-shot readout regime by incorporating a confusion matrix as the noise model (see Supplementary Section 2). Crucially, our approach builds upon the error-propagation method widely employed in single-parameter sensing and explicitly accounts for classical readout errors—a significant but frequently neglected factor in previous theoretical studies of quantum multiparameter estimation. By including these classical imperfections, we establish a comprehensive framework for quantitative multiparameter sensitivity analysis, providing practical insights and guidance for realistic implementations. In particular, when the classical readout noise dominates, optimal sensitivity occurs where the measurement signal slope is steepest.

In a standard Ramsey experiment for estimating the parameter $B_z$ in the Hamiltonian $H = \tfrac{B_z}{2}\sigma_z$, the steepest-slope condition can be realized either by introducing a bias field or by adjusting the measurement basis. This flexibility arises because the ultimate sensitivity bound, given by the quantum Fisher information (QFI) as $T^2$, is independent of $B_z$. This optimal sensitivity is achieved by initializing the system in an equal superposition state, such as $\ket{+}$, and performing measurements within the equatorial plane. Unfortunately, for multi-parameter sensing  applying a bias field is no longer a feasible strategy since $|\vec{B}| \neq 0$ generally leads to suboptimal sensitivity, as previously discussed. Instead, optimizing the measurement basis becomes the primary viable strategy. Specifically, we introduce a rotation gate \( U_r =\text{exp}\left(-i\frac{\pi}{3 \sqrt{3}}(\sigma^{e}_x+\sigma^{e}_y+\sigma^{e}_z)\right)\) prior to the disentangling gates. This rotation ensures uniform output probabilities $p_i = 1/4$ when starting from the initial state $\ket{\Phi_+}$, which is analogous to the optimal operating point $p = 1/2$ used in Ramsey experiments under classical noise. This strategy is close to the optimal basis rotation that provides a slight improvement over the uniform probability case, highlighting the intricate effects of classical noise in multiparameter estimation (see Supplementary Section 3).

\begin{figure*}
\centering
\includegraphics[width=\textwidth]{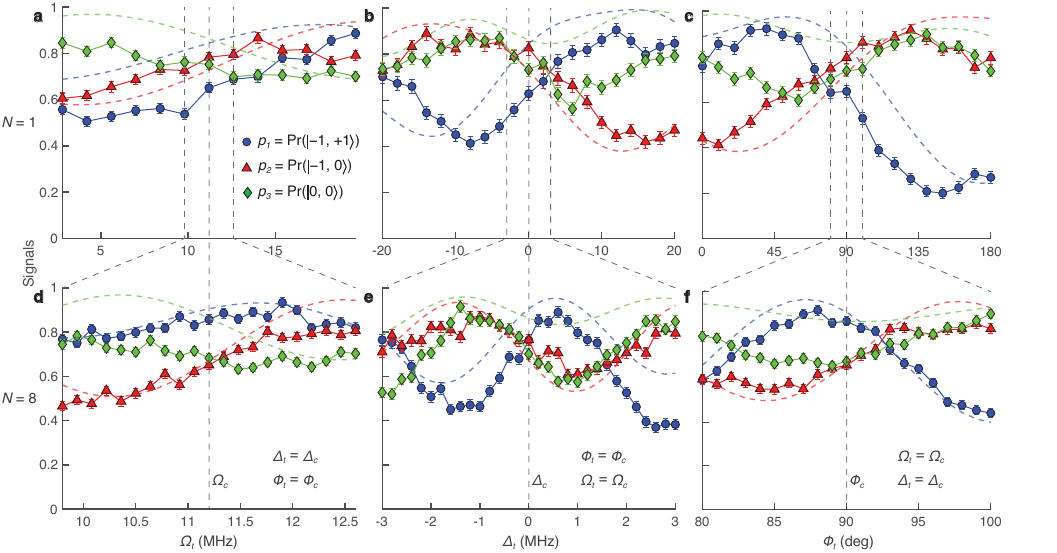}
\caption{\textbf{Experimental demonstration of multiparameter estimation.} \textbf{a}--\textbf{f}, Examples of measured signals as functions of the amplitude $\Omega_t$ (\textbf{a},\textbf{d}), frequency $\Delta_t$ (\textbf{b},\textbf{e}), and phase $\Phi_t$ (\textbf{c},\textbf{f}) of the target microwave field for two different numbers of repetitions, $N=1$ (\textbf{a}--\textbf{c}) and $N=8$ (\textbf{d}--\textbf{f}). Here, the control field parameters were set as follows: amplitude \(\Omega_c = (2\pi)\times11.2\ \text{MHz}\), frequency \(\Delta_c = 0\ \text{MHz}\), and phase \(\Phi_c = 90^\circ\). The sensitivity is limited by the uncertainty in signal measurement \(\sigma\approx0.02\), with the maximum slope point achieved at the parameters of the control field. At this point, the sensor is most sensitive to changes in the target parameters. 
Compared to the case of \(N = 1\), the signal is significantly more sensitive to variations in the target parameters when \(N = 8\). 
The dashed lines represent the model that accounts for SPAM errors due to laser irradiation, as shown in Fig.~\ref{fig:BSM}c. Error bars represent the
standard deviation of the signal. Each displayed point is a result of $n=3\times10^6$ averages of sequences.}
\label{fig:fig3}
\end{figure*}

\begin{figure}
\centering
\includegraphics[width=7cm]{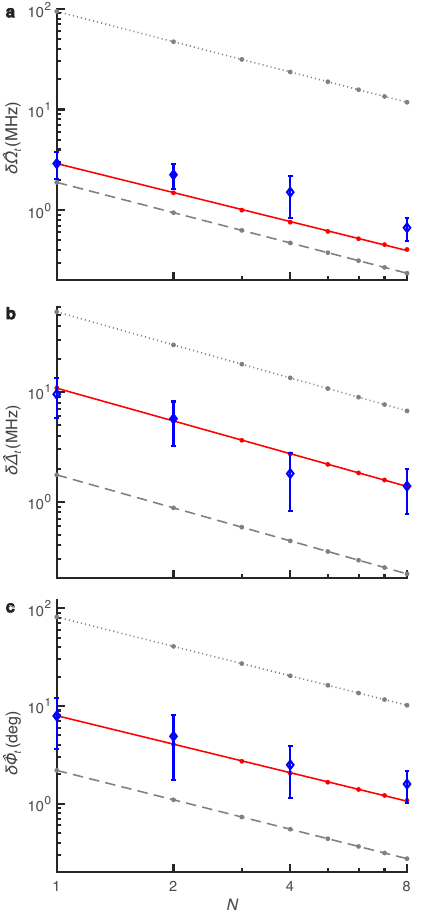}
\caption{\textbf{Characterization of sensitivity.} \textbf{a}--\textbf{c}, The minimum measurable amplitude (\textbf{a}), detuning (\textbf{b}), and phase (\textbf{c}) of the target microwave field as functions of number of repetitions $N$. The uncertainties in the estimates of the target parameters, calculated from error propagation, are plotted for both the simulation (red) and the experiment (blue) at the point of maximum sensitivity, i.e., around the parameters of the control field shown in Fig.~\ref{fig:fig3}. Error bars on the experimental data indicate the standard error of the uncertainty, obtained via error propagation from the standard error of the fitted slopes. A fit to the simulation (red line) shows that the sensitivity improves as the number of repetitions $N$, which is proportional to the interrogation time $T$, demonstrating the linear scaling for all three parameters. Gray dashed and dotted lines represent photon shot noise limited sensitivities for the ideal model with full and low ancilla polarization ($P = 1$ and $P = 0.51$), respectively. The divergence near $P = 0.5$ highlights the breakdown of simultaneous three-parameter estimation with a single qubit.}
\label{fig:fig4}
\end{figure}

To extract the experimentally determined Jacobian matrix, we sweep the target parameters and measure the resulting signal variations, as illustrated in Fig.~\ref{fig:fig3}. Each element of the Jacobian is obtained from the slope of the signal with respect to \((\Omega_t, \Delta_t, \Phi_t)\). Figure~\ref{fig:fig4} presents both measured and simulated sensitivities as a function of \(N = T/t\), where \(t = 30\,\text{ns}\) is fixed. The red solid lines are fits to $\delta \hat{\theta}_j\propto N^{-m}$ for $\theta_j \in \{\Omega_t,\Delta_t,\Phi_t\}$ for the simulation data (red circles), where $m\approx1$, which shows linear scaling with respect to the interrogation time. The experimental data, shown as blue diamonds, agree with these simulations. As seen in Fig.~\ref{fig:fig3}, increasing \(N\) makes the signals more sensitive to small variations in the target parameters, thereby illustrating why the sensitivity improves with \(N \propto T\).

Additionally, we plot sensitivities corresponding to two cases of the ideal vector field model: the photon shot noise limit and the case of low ancilla polarization. The gray dashed line represents the sensitivity in the ideal scenario with covariance matrix $\Sigma_{\hat{p}} = \sigma^2  \mathbf{I}$, defining the photon shot noise limit. In general, the photon shot noise limited sensitivity of NV sensors is typically about two orders of magnitude worse than the quantum projection limit, which, in our case, saturates the bound given by the QFIM. In Supplementary Section 3, we provide detailed comparisons between the photon shot noise limit and the quantum projection limit of the ideal model. The gray dotted line illustrates the photon shot noise limited sensitivity with ancilla initial polarization $P = 0.51$. As the polarization $P$ approaches 0.5, the sensitivity diverges, and at exactly $P = 0.5$, the Jacobian matrix becomes singular. This singularity corresponds to the fundamental limitation that three parameters cannot be simultaneously estimated with a single qubit. 

In addition to the aforementioned SPAM errors, several factors contribute to the degradation of sensitivity in our simulation model and actual experiments compared to the photon shot noise limit of the ideal model. These include pulse errors in sequential control, state preparation and measurement, finite pulse length effects of \(\pi\) pulses, and the presence of the frame change factor.

\section{\label{sec:level5}Conclusion and outlook}

Our demonstration achieved linear sensitivity scaling with respect to the interrogation time $T$ for three parameters, despite various experimental imperfections. In the ideal case of estimating a static vector field $\vec{B}$ with a sensor-ancilla system, the optimal sequential control scheme simultaneously achieves optimal precision for all three parameters, offering a threefold sensitivity improvement over separate single-parameter estimations with a single sensor~\cite{PhysRevLett.117.160801}.
It is worth noting that the target parameters in our case originate from a time-dependent AC field in the laboratory frame, specifically, its amplitude, frequency, and phase.
By applying a suitable rotating frame transformation and sequential controls, we effectively map the problem onto a time-independent Hamiltonian case, and achieve the scaling $\delta \hat{\theta}_j \propto 1/T$ for all parameters $\theta_j \in \{\Omega_t, \Delta_t, \Phi_t \}$. However, this does not necessarily correspond to the highest possible precision for each parameter, since the optimal scaling for estimating the frequency of an AC field can reach a quadratic scaling behavior~\cite{Pang2017,Schmitt2021}. In future work, we aim to investigate whether it is possible to simultaneously attain the highest precision for the parameters of the AC magnetic field \((\Omega,\Delta, \Phi)\) under realistic experimental conditions, and to explore potential trade-offs between these parameters.

 Additionally, when considering the total measurement time $T_{\text{tot}}$ as a resource, the time required for the Bell state preparation and the Bell state measurement, $T_{\text{m}}$, constitutes an overhead compared to the initialization and readout of single-parameter estimation schemes which usually require much simpler operations. However, if we suppose both the sensor and the ancilla enable single-shot readout, and the readout time dominates the whole measurement time ($T_{\text{tot}} \approx T_{\text{m}}$), the Bell state measurement—requiring only two readouts—might offer an advantage over performing three separate single-parameter estimations.

NV quantum sensors offer exceptional spatial resolution and versatility, making them particularly well-suited for multiparameter estimation in contexts where simultaneous measurement of multiple physical quantities at a single localized point is required. In such scenarios, the use of multiple bulky sensors is often impractical, further highlighting the advantages of NV-based approaches. In our demonstration, we focused on the simultaneous estimation of three parameters—the amplitude, detuning, and phase of a microwave drive. While microwave measurements have direct applications such as probing spin waves~\cite{10.1063/5.0175456,ogawa2024widebandwidefieldimagingspinwave,ogawa2025quantitativeimagingnonlinearspinwave}, in principle, our scheme can be extended to estimate any set of three parameters encoded in the Hamiltonian. Future research directions include the development of pulse sequences and Hamiltonian engineering techniques to simultaneously estimate multiple physical quantities accessible to the NV center,
such as vector components of static magnetic and electric fields, temperature, and pressure.

\begin{acknowledgments}
We thank Santiago Hern\'{a}ndez-G\'{o}mez, Yi-Xiang Liu, Yuichiro Matsuzaki, Naoki Yamamoto, Hiroshi Yano, and So Chigusa for insightful discussions.
This work was partially supported by the National Science Foundation under
Grant Nos.~PHY-1734011, MPS-2328774 and PHY-1734011 (the MIT-Harvard Center for Ultracold Atoms), by the National Research Foundation of Korea (NRF) via the Cooperative Research on Quantum Technology (2022M3K4A1094777),
and by the Research Grants Council of Hong Kong (14309223, 14309624, 14309022). T.~Isogawa acknowledges support from the Keio University Global Fellowship. 
\end{acknowledgments}

\bibliography{NV_multiparameter}

\newpage
\clearpage
\onecolumngrid
\vspace{2cm} 

\begin{center}
    {\Large \textbf{Supplementary Material: Entanglement-assisted multiparameter estimation with a solid-state quantum sensor}}\\[0.5cm]
\end{center}

\setcounter{section}{0}
\section{Ideal vector field model and mixed probe state}
\subsection{Ideal vector field model}

The task of estimating three parameters of an $\mathrm{SU}(2)$ unitary using a Bell state has been extensively studied in theory. We consider a two-level system subjected to a static magnetic field $\vec{B}$, parametrized in spherical coordinates as $\vec{B} = (B, \alpha, \beta)$, where $B$ denotes the field strength, $\alpha$ the polar angle, and $\beta$ the azimuthal angle. An ancillary system is assumed to remain uncoupled to $\vec{B}$~\cite{PhysRevLett.117.160801}. The system Hamiltonian is given by
\begin{equation}
    H(\vec{B}) = B \Bigl[\sin{(\alpha)} \,\cos{(\beta)} \,\sigma_x + \sin{(\alpha)}\,\sin{(\beta)}  \,\sigma_y + \cos{(\alpha)} \,\sigma_z \Bigr].\nonumber
\end{equation}
Let \(\rho_{SA}\) denote the initial state of the sensor-ancilla system. Its evolution under \(\vec{B}\) is described by
\begin{equation}
    \rho(\vec{B}) = U_A(\vec{B},T)\,\rho_{SA}\,U_A^\dagger(\vec{B},T),\nonumber
\end{equation}
where \(U_A(\vec{B},T) = e^{-i\,H(\vec{B})\,T}\otimes I_A\). One may then compute the quantum Fisher information matrix (QFIM) for \(\rho(\vec{B})\),
\begin{equation}
    \mathrm{QFI}_{i,j}(\rho(\vec B))=2\sum_{h,k} \frac{\bra{k}\partial_i\rho\ket{h}\bra{h}\partial_j\rho\ket{k}}{\bra{k}\rho\ket{k}+\bra{h}\rho\ket{h}}\nonumber,
    \label{eq:QFIm}
\end{equation}
where $\ket{h}$ are the eigenvectors of $\rho(\vec{B})$.

The QFIM can be maximized by choosing \(\rho_{SA}\) as any maximally entangled state. According to Ref.~\cite{PhysRevLett.117.160801}, this yields the optimal QFIM:
\begin{equation}
\mathrm{QFIM}^\mathrm{max}(\rho(\vec{B})) 
    = 
    4\begin{pmatrix}
    T^2 & 0 & 0\\
    0 & \sin^2\!\bigl(BT\bigr) & 0\\
    0 & 0 & \sin^2\!\bigl(BT\bigr)\,\sin^2\!\bigl(\alpha\bigr)
    \end{pmatrix}.\label{eqsup:QFIM1}
\end{equation}

A natural question is whether there exists a measurement (i.e., a POVM \(\{E_k\}\)) that saturates this bound, such that the classical Fisher information matrix (FIM) matches the optimal QFIM:
\begin{equation}
\mathrm{FIM}_{ij}(\rho(\vec B))=\sum_k\frac{\partial_i\Pr(E_k)\partial_i\Pr(E_k)}{\Pr(E_k)},\qquad
\Pr(E_k)=\textrm{Tr}[\rho(\vec B) E_k].\nonumber,
\end{equation}
In this case, such a measurement does exist. A projective measurement in the Bell basis gives the probabilities:
\begin{eqnarray}
p_1 &=& \cos^2\!\bigl(BT\bigr),\nonumber\\
p_2 &=& \sin^2\!\bigl(BT\bigr)\,\cos^2\!\bigl(\alpha\bigr),\nonumber\\
p_3 &=& \sin^2\!\bigl(BT\bigr)\,\sin^2\!\bigl(\alpha\bigr)\,\cos^2\!\bigl(\beta\bigr),\nonumber\\
p_4 &=& \sin^2\!\bigl(BT\bigr)\,\sin^2\!\bigl(\alpha\bigr)\,\sin^2\!\bigl(\beta\bigr)\nonumber,
\end{eqnarray}
which result in \(\mathrm{FIM} = \mathrm{QFIM}^\mathrm{max}(\rho(\vec{B}))\).

Now consider the sequential scheme illustrated in Fig.~\ref{fig:fig1}b of the main text. The total evolution is
\[
U_{FA}(\vec{B},Nt) = U_N U_A(\vec{B},t) \cdots U_2 U_A(\vec{B},t) U_1 U_A(\vec{B},t),
\]
where \( t = T/N \), and \( U_1, U_2, \ldots, U_N \) denote the sequential controls. 

Since \(\vec{B}\) is not known a priori, adaptive strategies based on estimates of \(\vec{B}\) are required. However, this does not alter the asymptotic scaling. Under an optimal feedback scheme, the QFIM becomes
\begin{equation}
\mathrm{QFIM}_N^{\max} \leq N^2 \mathrm{QFIM}_1^{\max} = 4N^2 
\begin{pmatrix}
t^2 & 0 & 0 \\
0 & \sin^2(Bt) & 0 \\
0 & 0 & \sin^2(Bt)\sin^2(\alpha)
\end{pmatrix}. \nonumber
\label{eqsup:QFIM}
\end{equation}

The upper bound is saturated asymptotically by choosing the controls as
\[
U_1 = U_2 = \cdots = U_N = U_A^\dagger(\vec{B} , t) = e^{iH(\vec{B} )t} \otimes I_A.
\]

In this case, a Bell state measurement again achieves the quantum Cramér-Rao bound.

To better interpret the estimation precision, we rewrite the Hamiltonian as
\[
H = B_x \sigma_x + B_y \sigma_y + B_z \sigma_z,
\]
with components \(B_x = B \sin\alpha \cos\beta\), \(B_y = B \sin\alpha \sin\beta\), and \(B_z = B \cos\alpha\). In the asymptotic limit, one can linearize the estimation around the true values:
\begin{align*}
\delta \hat{B}_x &= \sin \alpha \cos \beta \, \delta \hat{B} + B \cos \alpha \cos \beta \, \delta \hat{\alpha} - B \sin \alpha \sin \beta \, \delta \hat{\beta}, \\
\delta \hat{B}_y &= \sin \alpha \sin \beta \, \delta \hat{B} + B \cos \alpha \sin \beta \, \delta \hat{\alpha} + B \sin \alpha \cos \beta \, \delta \hat{\beta}, \\
\delta \hat{B}_z &= \cos \alpha \, \delta \hat{B} - B \sin \alpha \, \delta \hat{\alpha}.
\end{align*}

Squaring and summing yields the total variance:
\[
\delta \hat{B}_x^2 + \delta \hat{B}_y^2 + \delta \hat{B}_z^2 = \delta \hat{B}^2 + B^2 \delta \hat{\alpha}^2 + B^2 \sin^2(\alpha) \delta \hat{\beta}^2.
\]

This will be taken as the figure of merit for comparison (this is equivalent to considering a \textit{scalar} QFI, $\textrm{Tr}[\textrm{QFIM}\,W_a]$, with a weight matrix $W_a=\textrm{diag}[1,1/B^2,1/(B^2\sin^2(\alpha)]$.)

Under the optimal sequential scheme, we have
\begin{align}
\delta \hat{B}_x^2 + \delta \hat{B}_y^2 + \delta \hat{B}_z^2 &= \delta \hat{B}^2 + B^2 \delta \hat{\alpha}^2 + B^2 \sin^2(\alpha) \delta \hat{\beta}^2 \nonumber \\
&= \frac{1}{4nN^2} \left[ \frac{1}{t^2} + \frac{2B^2}{\sin^2(Bt)} \right], \nonumber
\end{align}
where $n$ is the number of repetitions. For a given \( T \), when \( N \to \infty \), \( t = T/N \to 0 \), \( [B^2/\sin^2(Bt)] \to (1/t^2) \), the precision limit under the optimal sequential feedback scheme reaches
\[
\delta \hat{B}_x^2 + \delta \hat{B}_y^2 + \delta \hat{B}_z^2 = \frac{3}{4nN^2 t^2} = \frac{3}{4nT^2}.
\]
For the estimation of an individual parameter \( B_i \), the optimal achievable precision is given by:
\[
\delta \hat{B}_i^2 = \frac{1}{4 n T^2}.
\]
To estimate three parameters sequentially—while keeping the total resources used constant—we assume that the measurement of each parameter is repeated \( n/3 \) times,
leading to a total precision for all three parameters:
\[
\delta \hat{B}_x^2 + \delta \hat{B}_y^2 + \delta \hat{B}_z^2 = \frac{3}{4 (n/3) T^2}= \frac{9}{4 n T^2},
\]
which aligns with the maximum precision achievable with \(N\) sensor qubits in parallel schemes~\cite{PhysRevLett.117.160801,PhysRevLett.116.030801}.

In the ideal model we described, the optimal sequential scheme can achieve the highest precision for all three parameters of a magnetic field simultaneously. We note that our experimental model  differs from the vector field model described above. In addition to the time-independent Hamiltonian dynamics \( U(\Omega, \Delta, \Phi) \), our system includes a frame change factor \( \exp(+i \Delta \sigma_z T / 2) \). This distinction is evident, for instance, from the fact that the total evolution \( U_t = \exp(+i \Delta \sigma_z T / 2) U(\Omega, \Delta, \Phi) \) cannot, in general, be written in a $T$-multiplicative form \( \exp(iGT)\), where \(G\) is an operator. 

\subsection{Mixed Probe State}\label{app:mixed}

As shown in the previous section, when using a maximally entangled state as the probe and performing a projective measurement in the Bell basis, the estimation can reach the local precision limit. However, experimental limitations might prevent the preparation of a perfect initial state. In this section, we consider the scenario where the measurement remains a Bell basis projection, but the probe state is mixed.
In particular, we assume an attempt to prepare an entangled state using Hadamard and CNOT gates as in the previous section, but starting from a state where the ancilla is mixed: $\rho_0 = \ket{0}\!\bra{0} \otimes (\mathbf{I} + P\,\sigma_z)/2$ instead of $\ket{00}$. This models the experimental limitation in polarizing the nuclear spin and broadly highlights the role of the ancilla qubit in achieving optimal metrological performance.

To evaluate the metrological bounds and compare the QFIM and the CFIM, we use scalar quantum and classical Fisher information metrics under two choices of weight matrices: $W = \mathbf{I}$ and $W_a = \mathrm{diag}[1, 1/B^2, 1/(B^2 \sin^2\alpha)]$. The latter corresponds to the mean squared error in estimating the components of the magnetic field vector.

In the limit $B \to 0$ (achievable via control and shown to be optimal in the pure-state case), the scalar quantum Fisher information becomes \[\mathrm{Tr}[\mathrm{QFIM}\,W_a] = T^2 (2 + P^2).\] Notably, if we assume no ancilla is used, the scalar QFI reduces to \[\mathrm{Tr}[\mathrm{QFIM}_1\,W_a] = 2T^2\] for a single qubit with an initial state $|+\rangle$.
We may express the QFI in terms of the concurrence $\mathcal{C}$, which quantifies entanglement. For our prepared state we have $\mathcal C=|P|$. Then, the scalar QFI becomes $\mathrm{Tr}[\mathrm{QFIM}\,W_a] = T^2 (2 + \mathcal{C}^2)$, highlighting that any gain over the ancilla-free scenario is due to the presence of entanglement.

The trace of the QFIM with the identity weight matrix, $\mathrm{Tr}[\mathrm{QFIM}]$, shows a more intricate angular dependence, even in the $B \to 0$ limit:
\[
\mathrm{Tr}[\mathrm{QFIM}] = T^2 [1 - (1 - P^2) \sin^2(\alpha) \cos^2(\beta)],
\]
while for the single-qubit case without ancilla,
\[
\mathrm{Tr}[\mathrm{QFIM}_1] = T^2 [1 - \sin^2(\alpha) \cos^2(\beta)].
\]
This shows that for $P \neq \pm1$, the QFI depends on the angles, reflecting no capability to simultaneously estimate all three parameters without an ancilla.
Importantly, in all these cases, even with a mixed ancilla, the QFI retains the $\propto T^2$ scaling. However the optimal scaling $\propto 3 T^2$ is only achieved for $P=\pm1$.

Lastly, we compare the quantum and classical Fisher information. While they may coincide for specific parameter values, in general, the trace of CFIM is substantially smaller than the quantum counterpart. This indicates that the Bell state measurement is no longer optimal when the initial state is mixed.

\section{Detailed Calculations of Sensitivity, Fisher Information, and Error Propagation}
\subsection{Quantum and Classical Noise}
In theoretical studies of quantum multiparameter estimation, sensitivity is typically assessed via the QFIM~\cite{Liu_2020}. In contrast, quantum sensor experiments often determine sensitivity through error-propagation methods~\cite{RevModPhys.89.035002}, which can readily incorporate classical readout noise arising from experimental constraints. Here, we generalize the single-parameter error-propagation approach to the multiparameter setting. This extension allows us to compare the resulting sensitivities with those derived from the QFIM or the CFIM. The crucial insight lies in moving from the binomial distribution in the single-parameter case to the multinomial distribution for the multiparameter case.

Quantum projection noise is one of the most fundamental sources of uncertainty in quantum sensing. In a projective measurement, rather than directly obtaining probabilities \(p_i\), one of \(K\) possible outputs is observed according to the multinoulli (or categorical) distribution. To accurately estimate \(p_i\), the experiment is repeated \(n\) times, and the number of occurrences for each output is collected into a histogram. The estimator for \(p_i\) is given by
\[
\hat{p}_i = \frac{n_i}{n},
\]
where \(n_i\) denotes the number of times outcome \(i\) is obtained, following a multinomial distribution. The total number of measurements is given by \(n = T_{\text{tot}}/(T + T_m),\) where \(T_{\text{tot}}\) is the total available measurement time, \(T\) is the measurement duration, and \(T_m\) is the additional time required for sensor initialization, manipulation, and readout~\cite{RevModPhys.89.035002}. The covariance matrix of \(p_i\) is therefore determined by that of the multinomial distribution,
\[
[\Sigma_{\hat{p},\text{quantum}}]_{i,j} = 
\begin{cases}
p_i (1-p_i)/n, & \text{if } i = j, \\[10pt]
- p_i p_j/n, & \text{if } i \neq j.
\end{cases}
\]
which indicates that projective measurements add noise on the order of \(1/\sqrt{n}\).

However, the dominant source of error is often the classical noise introduced during the readout of the sensor. Depending on whether the classical readout noise is small or large compared with the quantum projection noise, two regimes can be distinguished: the single-shot readout and the averaged readout. Because the efficiency of quantum state's readout is generally low, classical readout noise often becomes the dominant source of error. This is indeed the case for our experiments.

In the single-shot readout regime, the classical readout noise  is relatively low. The measurement apparatus produces a physical readout vector \(\vec{x}\), from which an outcome \(i\) is assigned. However, due to imperfections in the readout process, some of the \(\vec{x}_i\) values may be misclassified, thereby introducing a misclassification error. This error is described by the confusion matrix
\[
C_{ij} =
\begin{cases}
1 - \epsilon, & \text{if } i = j, \\[1mm]
\epsilon/(K-1), & \text{if } i \neq j,
\end{cases}
\]
where \(\epsilon\) is the misclassification rate and \(K\) is the number of possible outcomes. The effective probability \(\tilde{p}_i\) for outcome \(i\) is then given by
\[
\tilde{p}_i = \sum_{j=1}^{I} C_{ij} p_j = (1-\epsilon)p_i + \frac{\epsilon}{K-1}(1-p_i),
\]
which incorporates both  correctly-classified and misclassified events. The covariance matrix for the estimated probabilities, accounting for both quantum projection noise and the additional readout error, becomes
\[
[\Sigma_{\hat{p}}]_{ij} =
\begin{cases}
\displaystyle \tilde{p}_i (1-\tilde{p}_i)/n, & \text{if } i = j, \\[1mm]
\displaystyle -\tilde{p}_i \tilde{p}_j/n, & \text{if } i \neq j.
\end{cases}
\]
This formalism quantifies the effect of misclassification errors in the single-shot readout regime.

When classical noise during quantum state readout is significant, the estimate of \(p_i\) can be obtained from the average of the physical readout value \(y_i\):

\[
\hat{p}_i = \frac{\bar{y}_{p_i} - y_{p_i=0}}{y_{p_i=1} - y_{p_i=0}} ,
\]
where $y_{p_i=0}$ and $y_{p_i=1}$ denote the average physical readout values corresponding to $p_i=0$ and $p_i=1$, respectively.
The variance in the estimate of $p_i$ due to readout noise is given by 
\[
\sigma^2_{\hat{p}_i,\text{readout}} = \frac{\sigma_{y_i}^2}{(y_{p_i=1} - y_{p_i=0})^2}.
\]
In our NV center experiments, where photon shot noise dominates, the standard deviation of the readout can be approximated as \(\sigma_{y_{p_i}} \approx \sqrt{\bar{y}_{p_i}}\). The total covariance matrix of \(\hat{p}\) is then 
\[
\Sigma_{\hat{p}} = \Sigma_{\hat{p},\text{quantum}} + \text{diag}(\sigma^2_{\hat{p}_1,\text{readout}},\cdots,\sigma^2_{\hat{p}_K,\text{readout}}).
\]
In our case, since all \(\sigma^2_{\hat{p}_i,\text{readout}}\) are approximately equal and dominate the quantum contribution, i.e., \(\sigma^2_{\hat{p},\text{readout}} \gg [\Sigma_{\hat{p},\text{quantum}}]_{i,j}  \), we can approximate the total covariance as
\(
\Sigma_{\hat{p}} \approx \sigma^2_{\hat{p},\text{readout}}I.
\)

The covariance matrix for the estimate of the target parameters is obtained via error propagation:
\[
\Sigma_{\hat{\theta}} = J^{-1} \Sigma_{\hat{p}} \left(J^{-1}\right)^\intercal,
\]
where the Jacobian matrix \(J\) is defined as
\(
J_{i,j} = \partial p_i/\partial \theta_j.
\)
The square roots of the diagonal elements of \(\Sigma_{\hat{\theta}}\) yield the standard deviations of the estimates, providing a quantitative measure of sensitivity.

\subsection{Fisher Information and Error Propagation}

The Fisher information matrix quantifies the amount of information that an observable random variable carries about an unknown parameter. When the parameters of a probability distribution are reparameterized via a differentiable mapping, the corresponding Fisher information matrices are related through the Jacobian of the transformation—this is the essence of error propagation.

Let the probability distribution be parameterized by a vector \( p \) in one representation, and by an alternative vector \( \theta \) in another. The transformation between these parameterizations is given by a differentiable function, with the associated Jacobian matrix \( J \) defined as
\(
J_{ij} = \partial p_i/\partial \theta_j.
\)
Under this transformation, the Fisher information matrix in the \( \theta \) representation is given by
\[
\text{FIM}(\theta) = J^\top \, \text{FIM}(p) \, J.
\]
For an efficient estimator, which saturates the Cramér–Rao bound, the covariance matrix of the estimator is the inverse of the Fisher information matrix. That is,
\[
\Sigma_{\hat{\theta}} = \bigl(\text{FIM}(\theta)\bigr)^{-1}, \qquad
\Sigma_{\hat{p}} = \bigl(\text{FIM}(p)\bigr)^{-1}.
\]
Combining these expressions with the transformation rule for the Fisher information yields a relation between the covariance matrices in the two parameterizations:
\[
\Sigma_{\hat{\theta}} = J^{-1} \, \Sigma_{\hat{p}} \, (J^{-1})^\top.
\]
This framework illustrates that, under a change of variables, the Jacobian governs not only the transformation of the Fisher information matrix but also ensures that the associated error bounds for efficient estimators remain consistent across parameter spaces.

\section{Detailed sensitivity analysis on ideal vector field model}
\begin{figure}[htbp]
\centering
\includegraphics[width=0.75\columnwidth]{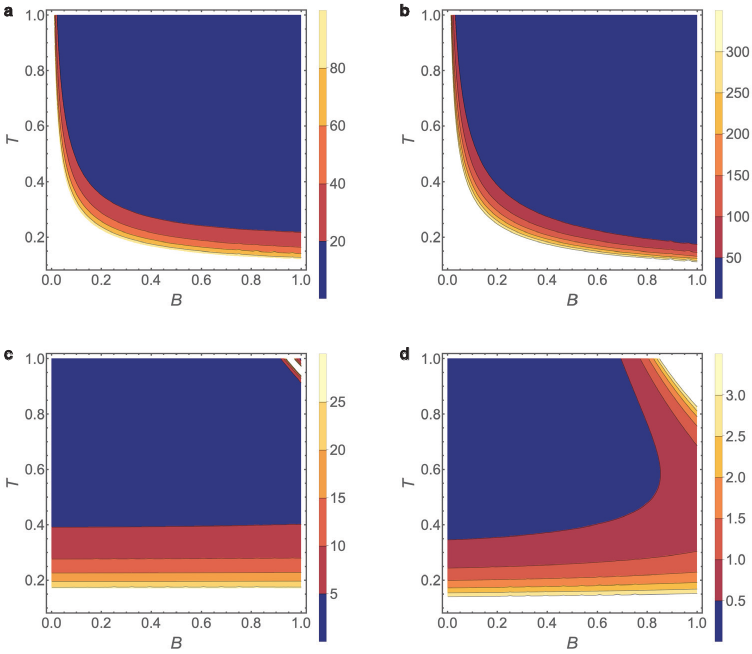}
\caption{Contour plots of the figure of merit $\delta \hat{B}_x^2 + \delta \hat{B}_y^2 + \delta \hat{B}_z^2 $. Panel (a) displays the single-shot case for the unrotated measurement basis, while panel (b) shows the result for the averaged readout case. Panels (c) and (d) present the analogous plots for the measurement basis rotated by \(
U = \exp\!\Bigl(-i\frac{\pi}{3 \sqrt{3}} (\sigma_x+\sigma_y+\sigma_z)\Bigr),
\). In all cases, the plots are functions of the magnetic field $B$ and the evolution time $T$, with the angular parameters fixed at $\alpha=\pi/4$ and $\beta=\pi/4$. Parameters are set as $\epsilon=0.01$, $n=1$ for (a) and (c), and $\sigma=0.01$ for (b) and (d).}
\label{fig:ErrorPropIdeal}
\end{figure}
\subsection{Quantum projection noise}

We begin with the ideal case in the absence of readout errors. The four outcome probabilities are given by:
\begin{eqnarray}
p_1&=&\cos^2{(BT)},\nonumber\\
p_2&=&\sin^2{(BT)}\cos^2{(\alpha)},\nonumber\\
p_3&=&\sin^2{(BT)}\sin^2{(\alpha)}\cos^2{(\beta)},\nonumber\\
p_4&=&\sin^2{(BT)}\sin^2{(\alpha)}\sin^2{(\beta)}.\nonumber
\end{eqnarray}
Since the probabilities must sum to one, $p_4$ can be written as $p_4=1-p_1-p_2-p_3$, allowing us to reduce the problem to three independent parameters. Note that in the single-parameter case, we similarly avoid using the $2\times2$ covariance matrix due to redundancy. The covariance matrix for the estimates of the probabilities $(p_1,p_2,p_3)$ is:
\[
\Sigma_{\hat{p}} =\frac{1}{n}
\begin{pmatrix}
p_1 (1 - p_1) & -p_1 p_2 & -p_1 p_3  \\
-p_2 p_1 & p_2 (1 - p_2) & -p_2 p_3 \\
-p_3 p_1 & -p_3 p_2 & p_3 (1 - p_3)  
\end{pmatrix},
\]
where $n$ is the number of measurements (or independent sensors.)
Using error propagation, the covariance matrix for the estimated parameters is given by:
\[
\Sigma_{\vec{B}} = J^{-1}\,\Sigma_{\hat{p}}\,(J^{-1})^\intercal
=
\frac{1}{4n}\begin{pmatrix}
\frac{1}{T^2} & 0 & 0 \\
0 &  \csc^2(B T) & 0 \\
0 & 0 &  \csc^2(B T) \csc^2(\alpha)
\end{pmatrix},
\]
which is consistent with the optimal bound given by the QFIM~\cite{PhysRevLett.117.160801}, see Eq.~\ref{eqsup:QFIM1}.

\subsection{Classical Readout Error: Single-Shot Readout}

We now consider the effect of classical readout errors in the single-shot measurement regime. We assume symmetric misclassification errors: if the true category is \( j \), the measurement outcome is correctly assigned with probability \( 1 - \epsilon \), and incorrectly assigned to each of the other three categories with equal probability \( \epsilon / 3 \).
Given the true probabilities \( p_j \), the observed probabilities \( q_j \) are then:
\[
q_j = p_j (1 - \epsilon) + \frac{\epsilon}{3} \sum_{i \neq j} p_i = p_j \left(1 - \frac{4\epsilon}{3}\right) + \epsilon.
\]

The covariance matrix for the estimates of the observed outcomes \(q = (q_1, q_2, q_3)\) is:
\[
\Sigma_{\hat{q}} = \frac{1}{n}\begin{pmatrix}
q_1 (1 - q_1) & -q_1 q_2 & -q_1 q_3  \\
-q_2 q_1 & q_2 (1 - q_2) & -q_2 q_3 \\
-q_3 q_1 & -q_3 q_2 & q_3 (1 - q_3)
\end{pmatrix}.
\]
The corresponding parameter covariance matrix is obtained via error propagation:
\(
\Sigma_{\vec{B}} = J^{-1} \Sigma_{\hat{q}} \left(J^{-1}\right)^\intercal.
\)
We again compute the figure of merit
\(
\delta \hat{B}_x^2 + \delta \hat{B}_y^2 + \delta \hat{B}_z^2 .
\)
Figure~\ref{fig:ErrorPropIdeal}(a) shows the result. Notably, at $B=0$, which is the optimal point in the absence of noise, the value diverges and the sensitivity becomes the worst.

\subsection{Classical Readout Error: Averaged Readout}

Here, we focus on the scenario where readout error dominates, a condition that is particularly relevant to our experiment. In this case,  the measurement noise is assumed to be isotropic and uncorrelated, so that
\(
\Sigma_{\hat{p}} = \sigma^2 \mathbf{I}.
\)

Using error propagation, we obtain
\(
\Sigma_{\vec{B}} = J^{-1}\,\Sigma_{\hat{p}}\,\left(J^{-1}\right)^\intercal.
\)

Explicitly, this gives:
\begin{align}
\Sigma_{\vec{B}} = \scalebox{0.75}{$
\begin{pmatrix}
\frac{\sigma^2\, \csc^2(2 B T)}{T^2} & \frac{\sigma^2\, \cot(\alpha)\, \csc^3(B T)\, \sec(B T)}{4 T} & \frac{\sigma^2\, \cot(\beta)\, \csc^3(B T)\, \csc^2(\alpha)\, \sec(B T)}{4 T} \\[10pt]
\frac{\sigma^2\, \cot(\alpha)\, \csc^3(B T)\, \sec(B T)}{4 T} & \frac{\sigma^2}{32}\left( 11 + 4\cos(2\alpha) + \cos(4\alpha) \right) \csc^4(B T)\, \csc^2(\alpha)\, \sec^2(\alpha) & \frac{\sigma^2}{8}\left( 3 + \cos(2\alpha) \right) \cot(\beta)\, \csc^4(B T)\, \csc^3(\alpha)\, \sec(\alpha) \\[10pt]
\frac{\sigma^2\, \cot(\beta)\, \csc^3(B T)\, \csc^2(\alpha)\, \sec(B T)}{4 T} & \frac{\sigma^2}{8}\left( 3 + \cos(2\alpha) \right) \cot(\beta)\, \csc^4(B T)\, \csc^3(\alpha)\, \sec(\alpha) & \frac{\sigma^2}{16}\left( 7 + 4\cos(2\beta) + \cos(4\beta) \right) \csc^4(B T)\, \csc^4(\alpha)\, \csc^2(\beta)\, \sec^2(\beta)
\end{pmatrix}
$}\nonumber
\end{align}

The corresponding figure of merit becomes:
\begin{align}
\delta \hat{B}_x^2 + \delta \hat{B}_y^2 + \delta \hat{B}_z^2  &= \delta \hat{B}^2 + B^2 \delta \hat{\alpha}^2 + B^2 \sin^2(\alpha) \delta \hat{\beta}^2\nonumber \\
&= \frac{\sigma^2\, \csc^2(B T)}{4 T^2} \left[ \sec^2(B T) + B^2 T^2\, \csc^2(B T) \left( \csc^2(\alpha)\left( 3 \csc^2(\beta) + \sec^2(\beta) \right) + \tan^2(\alpha) \right) \right].\nonumber
\end{align}

As shown in Fig.~\ref{fig:ErrorPropIdeal}(b), the result again diverges at \( B = 0 \), and the sensitivity is the worst.

\subsection{Rotation of the Bell Measurement Basis}

Interestingly, the loss of sensitivity at \( B = 0 \) due to classical noise can be recoverd by rotating the measurement basis. In particular, we apply the unitary transformation:
\(
U = \exp\left(-i\frac{\pi}{3 \sqrt{3}}(\sigma_x + \sigma_y + \sigma_z)\right),
\)
which redefines the measurement basis.

In this rotated basis, the outcome probabilities become
\begin{align}
p_1 &= \frac{1}{4} \left[\cos(BT) + \sin(BT)\Bigl(\cos(\alpha) + \sin(\alpha)\bigl(\cos(\beta) + \sin(\beta)\bigr)\Bigr)\right]^2, \nonumber\\
p_2 &= \frac{1}{4} \left[\cos(BT) - \sin(BT)\Bigl(\cos(\alpha) + \sin(\alpha)\bigl(\cos(\beta) - \sin(\beta)\bigr)\Bigr)\right]^2, \nonumber\\
p_3 &= \frac{1}{4} \left[\cos(BT) + \sin(BT)\Bigl(\cos(\alpha) - \sin(\alpha)\bigl(\cos(\beta) + \sin(\beta)\bigr)\Bigr)\right]^2, \nonumber\\
p_4 &= \frac{1}{4} \left[\cos(BT) - \sin(BT)\Bigl(\cos(\alpha) + \sin(\alpha)\bigl(-\cos(\beta) + \sin(\beta)\bigr)\Bigr)\right]^2.\nonumber
\end{align}
At \( B = 0 \), all four probabilities become equal: \( p_1 = p_2 = p_3 = p_4 = 1/4 \).
With these probabilities, one finds again
\[
\Sigma_{\vec{B}} = J^{-1}\,\Sigma_{\hat{p}}\,(J^{-1})^\intercal
= \frac{1}{4n}\begin{pmatrix}
\frac{1}{T^2} & 0 & 0 \\
0 &  \csc^2(BT) & 0 \\
0 & 0 &  \csc^2(BT)\,\csc^2(\alpha)
\end{pmatrix}
\]
for the noiseless case.

We perform error propagation for both the single-shot and averaged readout cases to obtain the covariance matrix $\Sigma_{\vec{B}}$. In general, concise expressions for the covariance matrix and the figure of merit, \(
\delta \hat{B}_x^2 + \delta \hat{B}_y^2 + \delta \hat{B}_z^2 \) are not available. Instead, we numerically evaluate the figure of merit as a function of \(B\) and \(T\), as shown in Fig.~\ref{fig:ErrorPropIdeal}~c and d for the two readout schemes. In both cases, the optimal point at $B=0$ is recovered.

\subsection{Comparison of multiparameter estimation and repeated single-parameter estimation under averaged readout}

We now compare the sensitivity at the optimal point $B=0$, recovered by the aforementioned basis rotation in the averaged readout scenario, with the sensitivity achievable via single-parameter estimation. For averaged readout, the optimal variance for each component $B_i$ is given by $\sigma^2/T^2=\sigma_0^2/nT^2$, where $\sigma_0^2$ is the variance of each repetition. To estimate three parameters sequentially under a fixed total resource constraint, we assume that each parameter is measured independently and repeated \( n/3 \) times, yielding a figure of merit:
\[
\delta \hat{B}_x^2 + \delta \hat{B}_y^2 + \delta \hat{B}_z^2 = \frac{3\sigma_0^2}{(n/3)T^2} = \frac{9\sigma_0^2}{nT^2}.
\]
In contrast, for the multiparameter estimation scenario with $\Sigma_{\hat{p}} = (\sigma_0/\sqrt{n})^2 \mathbf{I}$, the corresponding figure of merit is
\begin{align*}
\delta \hat{B}_x^2 + \delta \hat{B}_y^2 + \delta \hat{B}_z^2&=
\mathrm{tr}(\Sigma_{\vec{B}}) \\
&= (\sigma_0/\sqrt{n})^2 \mathrm{tr}\left(J^{-1}(J^{-1})^{\mathsf{T}}\right).
\end{align*}
The Jacobian $J$ at the point $B_x=B_y=B_z=0$ can be evaluated analytically as
\begin{align}
J = \frac{T}{2}
\begin{pmatrix}
1 & \;\;1 & \;\;1 \\
-1 & \;\;1 & -1 \\
-1 & -1 & \;\;1
\end{pmatrix}.\nonumber
\end{align}
Using this Jacobian, the figure of merit for the multiparameter scheme becomes
\begin{align}
\delta \hat{B}_x^2 + \delta \hat{B}_y^2 + \delta \hat{B}_z^2 = \frac{6\sigma_0^2}{nT^2}.\nonumber
\end{align}

Thus, the multiparameter estimation scheme demonstrates a sensitivity advantage over repeated single-parameter estimation even in the case of averaged readout.

\subsection{Exact Optimization for Averaged Readout}

We present here the exact solution for sensitivity optimization through a rotation of the Bell basis in the case of averaged readout. The rotation operator is given by:
\[
\exp\left(-i\frac{\pi c}{2}\left[\cos(a)\sigma_z + \sin(a)\cos(b)\sigma_x + \sin(a)\sin(b)\sigma_y\right]\right).
\]

The resulting probabilities for measurement outcomes are:

\begin{align}
p_1 &= \left[\sin\left(\frac{\pi c}{2}\right)\sin(B T)\left(\sin(a)\sin(\alpha)\cos(b-\beta)+\cos(a)\cos(\alpha)\right)+\cos\left(\frac{\pi c}{2}\right)\cos(B T)\right]^2,\nonumber\\[10pt]
p_2 &= \left[\cos\left(\frac{\pi c}{2}\right)\sin(\alpha)\cos(\beta)\sin(B T) - \sin\left(\frac{\pi c}{2}\right)\left(\sin(B T)\left(\sin(a)\sin(b)\cos(\alpha)-\cos(a)\sin(\alpha)\sin(\beta)\right) + \sin(a)\cos(b)\cos(B T)\right)\right]^2,\nonumber\\[10pt]
p_3 &= \left[\sin(B T)\left(\sin\left(\frac{\pi c}{2}\right)\sin(a)\sin(\alpha)\sin(b-\beta)+\cos\left(\frac{\pi c}{2}\right)\cos(\alpha)\right)-\sin\left(\frac{\pi c}{2}\right)\cos(a)\cos(B T)\right]^2.\nonumber
\end{align}

Using these probabilities, we compute the Jacobian matrix with respect to $(B,\alpha,\beta)$. Assuming uniform noise for the outcomes \((\sigma_0/\sqrt{n})^2\mathbf{I}\) and weight matrix \(W=\{1,B^2,B^2\sin^2\alpha\}\), we calculate the figure of merit:
\[
\delta \hat{B}_x^2 + \delta \hat{B}_y^2 + \delta \hat{B}_z^2 =\mathrm{Tr}\left[W J^{-1}(J^T)^{-1}\left(\frac{\sigma_0^2}{n}\right)\right].
\]

In the limit \(B \to 0\) (which corresponds to applying optimal control to recover the bias point), this simplifies to:
\[
\delta \hat{B}_x^2 + \delta \hat{B}_y^2 + \delta \hat{B}_z^2
  \;\to\;
  \frac{\sigma_0^2\csc^2\!\left(\frac{\pi c}{2}\right)
        \Bigl[\csc^2(a)\!\bigl(3\csc^2(b)+\sec^2(b)\bigr)+\sec^2(a)\Bigr]
        +\sec^2\!\left(\frac{\pi c}{2}\right)}{4\,n\,T^2}.
\]

This expression achieves optimality at:
\[
a= \arctan \left(\sqrt{\sqrt{3}+1}\right), \quad b= \arctan\left(\sqrt[4]{3}\right), \quad \pi c = 2 \arctan \left(\sqrt{\sqrt{3}+2}\right),
\]
leading to the minimal value:
\[
\frac{3(\sqrt{3}+2)\sigma_0^2}{2nT^2}\approx \frac{5.598\,\sigma_0^2}{nT^2}.
\]

This result underscores the increased complexity of sensitivity optimization when classical noise is accounted for in quantum multiparameter estimation.

\subsection{Comparison between the quantum projection limit and the photon shot noise limit}
Here, we compare the quantum projection limit and the photon shot noise limit within the ideal vector field model discussed in the main text. The quantum projection limit refers to the sensitivity attainable when considering only quantum projection noise. Under these ideal conditions and employing optimal sequential control strategies, the quantum projection limit is directly bounded by the quantum Fisher information matrix presented in Supplementary Section 1. Figure \ref{fig:projection_vs_shot} contrasts these sensitivity limits, emphasizing how measurement noise during the detection stage impacts the achievable sensitivity.

\begin{figure}[htbp]
\centering
\includegraphics[width=\columnwidth]{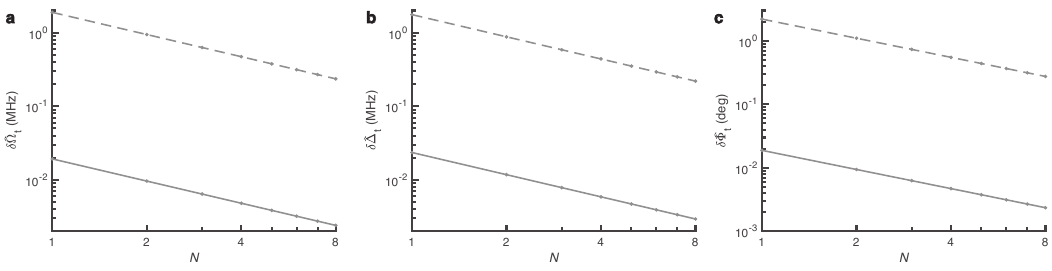}
\caption{Comparison of sensitivities between the quantum projection limit and the photon shot noise limit for the ideal vector field model as a function of the number of sequential control steps \(N\). Panels (a--c) show sensitivities for estimating parameters of the evolution operator \(
U_t = \exp\left[-i \left(\frac{\Delta_t}{2} \sigma^{e}_z + \frac{\Omega_t}{2} (\cos{\Phi_t} \, \sigma^{e}_x - \sin{\Phi_t} \, \sigma^{e}_y) \right) T \right].
\) Solid lines represent the quantum projection limit, computed from the covariance matrix of a multinomial distribution, representing ideal conditions where sensitivity is limited only by quantum projection noise. Dashed gray lines represent sensitivities under photon shot noise, calculated using the covariance matrix \(\sigma^2 \mathbf{I}\), and correspond to those shown in Fig.~\ref{fig:fig4} of the main text. This figure highlights the fundamental sensitivity gap due to measurement noise at the final readout stage. The parameters used here match those in Fig.~\ref{fig:fig4} of the main text.}
\label{fig:projection_vs_shot}
\end{figure}

\section{Quantitative evaluation of the effect of laser irradiation on the nitrogen nuclear spin state}\label{app:laser}

\begin{figure}[t]
    \centering

    \includegraphics[width=0.4\columnwidth]{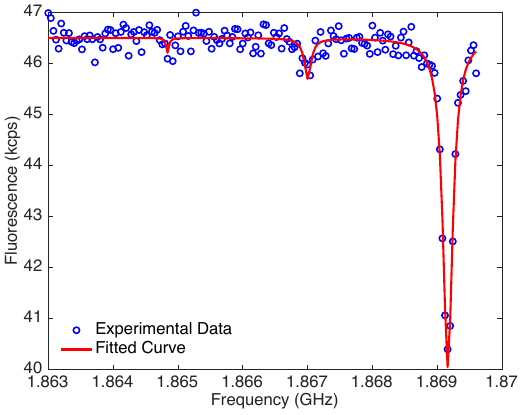}
    \caption{Pulsed ODMR measurement as a function of the microwave frequency (GHz). 
    A triple-Lorentzian model (red line) is fitted to the data, identifying three resonance features around 1.87\,GHz. The fit parameters include the amplitudes, widths, and center frequencies of each Lorentzian peak, as well as a global offset. The dimensionless ratio
    \(
    \texttt{P}_{+1}/(\,\texttt{P}_{+1} + \texttt{P}_{0} + \texttt{P}_{-1})\,
    \)
    represents the polarization rate \(P\), where each \(\texttt{P}_{i},\,i\in\{-1,0,+1\}\) represents the height of the Lorenzian peak.}
    \label{fig:triple_lorentzian}
\end{figure}

\begin{figure}[t]
    \centering
    \includegraphics[width=\columnwidth]{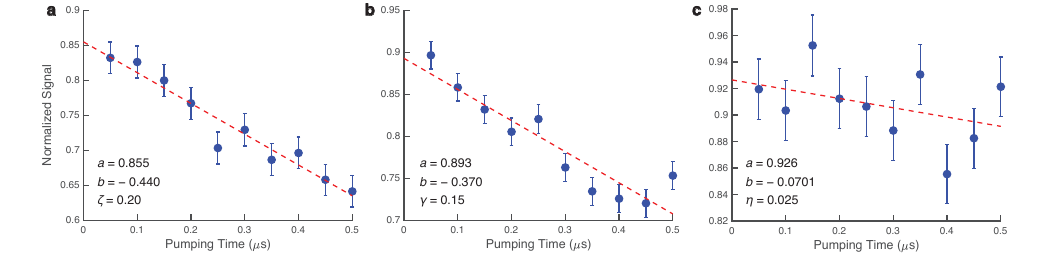}
    \caption{Normalized signal measurements with varying pumping times. In each panel, blue markers with error bars represent the normalized experimental data, while the red dashed line indicates the linear regression fit. Error bars represent the
standard deviation of the signal. Panel (a) corresponds to the $\zeta$, panel (b) to the $\gamma$, and panel (c) to the $\eta$.}
    \label{fig:leakage_rate}
\end{figure}

Although laser illumination does not directly act on the nuclear spin, hyperfine coupling to the NV electronic spin can modify the nuclear spin state during optical pumping. In our experiments, we therefore select a bias field \(B^0_z= 357\,\mathrm{G}\) that balances two competing needs: effective nuclear spin polarization and an electronic spin readout insensitive to the nuclear spin state. When \(B^0_z\) is set close to the excited-state level anti-crossing (ESLAC)~\cite{PhysRevLett.102.057403}, a long \(\sim20~\mu\)s green laser pulse optically pumps the nuclear spin into the \(m_I=+1\) state. We measured the polarization of the \( m_S = 0, m_I = +1 \) state after a 20~\(\mu\)s green laser pulse using pulsed ODMR and obtained a value of \( P = 0.85 \) (Fig.~\ref{fig:triple_lorentzian}). Additionally, we measure the effect on nuclear spin states following laser irradiation of up to 500~ns.
Specifically, we first prepare the states \( m_S = 0, m_I = +1,0,-1 \) and then measure the nuclear spin states after applying laser pulses with durations ranging from 0~ns to 500~ns. Note that the allowed transitions in this case are as follows:
\begin{enumerate}
    \item \( m_S = 0, m_I = 0 \rightarrow m_S = 0, m_I = +1 \) (leakage rate: \(\zeta \))
    \item \( m_S = 0, m_I = -1 \rightarrow m_S = 0, m_I = +1 \) (leakage rate: \( \gamma \))
    \item \( m_S = 0, m_I = -1 \rightarrow m_S = 0, m_I = 0 \) (leakage rate: \( \eta \))
\end{enumerate}

As illustrated in Fig.~\ref{fig:leakage_rate}, by fitting these experimental data with a linear model, we define the leakage rates (\( \zeta \), \( \gamma \), and \( \eta \)) corresponding to each transition after 400~ns of laser irradiation.

\section{Modeling of SPAM Errors Due to Non-Ideal Behavior of Nuclear Spin States Under Laser Irradiation}
\subsection{State preparation}
The pulse sequence for creating the Bell state, as shown in Figure 1, is straightforward: a \(\pi/2\) pulse is applied to the nuclear spin, followed by a selective \(\pi\) pulse on the electronic spin. However, imperfect polarization leads to an imperfect initial state.  At the chosen static magnetic field strength, while the electronic spin is nearly fully polarized by a long laser pulse, the nuclear spin remains only partially polarized, with a polarization of \( P = 0.85 \). Consequently, the initial state is given by the mixed state \(\rho = P\,|0,+1\rangle\langle0,+1| + (1-P)\,|0,0\rangle\langle0,0|\), which evolves under the entangling gates to \(\rho = P | \Phi_+ \rangle \langle \Phi_+ | + (1 - P) | \Phi_- \rangle \langle \Phi_- |,\) where \(|\Phi_{\pm}\rangle=(|0,0\rangle \pm |-1,+1\rangle)/\sqrt{2}\) are two of the Bell states. Compared to the ideal case of \( P = 1 \), when \( P < 1 \), the signal amplitude decreases, reducing sensitivity, as already discussed in Sec.\ref{app:mixed}.

\subsection{Measurement}
We analyze the effect of nuclear spin state leakage on the measurement results when reading out electronic spins with short laser pulses with a simple and intuitive linear model based on the leakage rates defined in Supplementary Section~\ref{app:laser}, although the nuclear spin dynamics under laser irradiation could be rigorously described by the Lindblad master equation. 

Let us denote the populations before the readout sequence as 
\[
p_1 = \Pr(\lvert -1, +1 \rangle),\quad
p_2 = \Pr(\lvert -1, 0 \rangle),\quad
p_3 = \Pr(\lvert 0, 0 \rangle),\quad
p_4 = \Pr(\lvert 0, +1 \rangle).
\]
Note that the measured result of $p_1$ is unaffected by leakage. However, after reading out $p_1$ with the first short laser pulse, the nuclear spin states with $m_s = 0$ become 
\[
\Pr(\lvert 0,0\rangle) = (1-\zeta)\,p_2 + \eta\,p_3, 
\qquad
\Pr(\lvert 0,-1\rangle) = (1-\gamma-\eta)\,p_3
\]
due to leakage. Consequently, the population read out by the next short laser pulse is 
\[
(1-\zeta)\,p_1 + \eta\,p_3.
\]
During this interval, the state $\Pr(\lvert 0,-1\rangle)$ further evolves to 
\[
\Pr(\lvert 0,-1\rangle) = (1-2\gamma - 2\eta)\,p_3,
\]
and this final value is read out by the last short laser pulse.

In summary, the effect of leakage can be expressed by the following linear transformation:
\[
\begin{aligned}
p_1' &= p_1,\\
p_2' &= (1-\zeta)\,p_2 + \eta\,p_3,\\
p_3' &= (1 - 2\gamma - 2\eta)\,p_3.
\end{aligned}
\]
We used these expressions to simulate the expected experimental results and compared them with the data in the main text. 

\section{Dynamics and control of the two-qubit system}
The NV center provides a spin system composed of an electronic spin ($S = 1$) and a nitrogen nuclear spin ($I = 1$), which interact strongly through hyperfine coupling ($A \approx -(2\pi)\times2.16$ MHz). The hyperfine interaction induces a frequency shift, allowing selective microwave pulse by tuning the transition rate $\Omega_e$. The radio frequency is on resonance with the transition between $|0,+1\rangle$ and $|0,0\rangle$. The transition rate is about $\Omega_n = (2\pi)\times20-40$ kHz. Because the nuclear Rabi transition rate is always lower than the detuning given by the hyperfine interaction ($\Omega_n \ll A$), only selective driving of the ancillary spin is available and unconditional gates must be engineered with composite pulses.

\subsection{Nuclear Spin $\pi$ and $\pi/2$ Pulses}

A single radio-frequency (RF) pulse suffices to implement the initial \(\pi/2\) gate in Fig.~1 of the main text, because the electronic spin is initially in the \(m_S=0\) state. However, the final \(\pi/2\) gate must be non-selective, and we realize it by applying two selective \(\pi/2\) pulses separated by a \(\pi\) pulse on the electronic spin. During each selective pulse, the \(m_S=-1\) manifold continues to evolve under the hyperfine interaction. To cancel this evolution, we set the duration of the RF \(\pi/2\) pulse to an integer multiple of \(2\pi / A\). Furthermore, we compensate for the phase of the second \(\pi/2\) pulse applied to the nuclear spin by adjusting it by \( A\tau/2 \), where \(\tau\) is the time interval between the first and second \(\pi/2\) pulses on the nuclear spin. This correction effectively cancels the time evolution induced by the \(\sigma_z^{n}\) term in \( H_{\text{int}} \) during that period.

\subsection{Electronic Spin $\pi$ and $\pi/2$ Pulses}

A strong driving $\Omega_e \approx(2\pi)\times 25$ MHz $\gg A$, is used to drive the qubit non-selectively for any ancilla state, whereas $\Omega_e = (2\pi)\times1.25$ MHz $< A$ is used to drive the qubit selectively on the ancilla state, thus engineering conditional gates. In our experiments, we aim to drive a $\pi$ pulse on one hyperfine transition (corresponding to $m_I = +1$) while simultaneously ensuring that the neighboring transition (corresponding to $m_I = 0$) effectively undergoes a $2\pi$ rotation and thus remains unaffected. By implementing this scheme, we can reduce unwanted population transfer on the off-resonant transition. Let $\Omega_{0}$ be the on-resonance Rabi frequency. Out of resonance, where the detuning $\delta$ is equal to the hyperfine splitting $A= 2.16$ MHz, the generalized Rabi frequency is
$\Omega_{\mathrm{hf}} \;=\; \sqrt{\Omega_{0}^2 + \delta^2}.$
We require that the off-resonant drive perform a $2\pi$ rotation at the same time the on-resonance drive performs a $\pi$ rotation, hence $2\,\Omega_{0} \;=\; \Omega_{\mathrm{hf}} \;=\; \sqrt{\Omega_{0}^2 + \delta^2}.$
Solving for $\Omega_{0}$ gives $\Omega_{0} \;=\; \delta/\sqrt{3}\;\approx\; 1.25$ MHz.
Hence, the resonant $\pi$ pulse duration is $t_{\pi} \;=\; \pi/\Omega_{0} \;\approx\; 0.40\,\mu$s. Under these conditions, the neighboring transition experiences a $2\pi$ pulse and remains effectively unperturbed, thereby improving the gate fidelity.

\end{document}